# A negative permeability material at red light

Hsiao-Kuan Yuan, Uday K. Chettiar, Wenshan Cai, Alexander V. Kildishev,
Alexandra Boltasseva*, Vladimir P. Drachev, and Vladimir M. Shalaev

*Birck Nanotechnology Center, Purdue University, West Lafayette, IN 47907, USA*
*\* on leave from DTU, Research Center COM and Nanophotonics, DK-2800 Kgs. Lyngby, Denmark*
*kildishev@purdue.edu*

**Abstract:** Experimental demonstration of a negative permeability in a periodic array of pairs of thin silver strips is presented. Two samples with different strip thicknesses are fabricated; optical measurements of the samples confirm our initial design projections by showing the real part of permeability to be about −1 for the sample with thinner strips and −1.7 for the sample with thicker strips at wavelengths of 770 nm and 725 nm, respectively.

**References and links**

1. D. R. Smith, S. Schultz, P. Markoŝ, and C. M. Soukoulis, "Determination of effective permittivity and permeability of metamaterials from reflection and transmission coefficients," Phys. Rev. B **65**, 195104 (2002).
2. A. V. Kildishev, W. Cai, U. K. Chettiar, H.-K. Yuan, A. K. Sarychev, V. P. Drachev, and V. M. Shalaev, "Negative refractive index in optics of metal-dielectric composites," J. Opt. Soc. Am. B **23**, 423-433 (2006).
3. U. K. Chettiar, A. V. Kildishev, T. A. Klar, and V. M. Shalaev, "Negative index metamaterial combining magnetic resonators with metal films," Opt. Express **14**, 7872-7877 (2006).
4. G. Shvets and Y. A. Urzhumov, "Negative index meta-materials based on two-dimensional metallic structures," J. Opt. A **8**, S122 (2006).
5. V.A.Podolskiy, A.K. Sarychev, and V.M. Shalaev, "Plasmon modes in metal nanowires and left-handed materials," J. Nonlinear Opt. Phys. Materials **11**, 65-74 (2002).
6. S. Zhang, W. Fan, K. J. Malloy, S. R. J. Brueck, N. C. Panoiu, and R. M. Osgood, "Demonstration of metal-dielectric negative-index metamaterials with improved performance at optical frequencies," J. Opt. Soc. Am. B **23**, 434-438 (2006).
7. J. Zhou, L. Zhang, G. Tuttle, T. Koschny, and C. M. Soukoulis, "Negative index materials using simple short wire pairs," Phys. Rev. B **73**, 041101(R) (2006).
8. T. J. Yen, W. J. Padilla, N. Fang, D. C. Vier, D. R. Smith, J. B. Pendry, D. N. Basov, and X. Zhang, "Terahertz magnetic response from artificial materials," Science **303**, 1494-1496 (2004).
9. S. Linden, C. Enkrich, M. Wegener, J. Zhou, T. Koschny, and C. M. Soukoulis, "Magnetic response of metamaterials at 100 Terahertz," Science **306**, 1351-1353 (2004).
10. S. Zhang, W. Fan, B. K. Minhas, A. Frauenglass, K. J. Malloy, and S. R. J. Brueck, "Midinfrared resonant magnetic nanostructures exhibiting a negative permeability," Phys. Rev. Lett. **94**, 037402 (2005).
11. C. Enkrich, M. Wegener, S. Linden, S. Burger, L. Zschiedrich, F. Schmidt, J. F. Zhou, Th. Koschny, and C. M. Soukoulis, "Magnetic metamaterials at telecommunication and Visible frequencies," Phys. Rev. Lett. **95**, 203901 (2005).
12. A. N. Grigorenko, "Negative refractive index in artificial metamaterials," Opt. Lett. **31**, 2483-2485 (2006).
13. A. V. Kildishev, V. P. Drachev, U. K. Chettiar, D. Werner, D.-H. Kwon, and V. M. Shalaev, "Comment on "Negative Refractive Index in Artificial Metamaterials [A. N. Grigorenko, Opt. Lett., 31, 2483 (2006)]," http://arxiv.org/abs/physics/0609234, (2006).
14. A. V. Kildishev and U. K. Chettiar, "Cascading Optical Negative Index Metamaterials," submitted to the Journal of Applied Computational Electromagnetics Society.
15. P. B. Johnson and R. W. Christy, "Optical constants of the noble metals," Phys. Rev. B **6**, 4370-4379 (1972).
16. G. Dolling, M. Wegener, C. M. Soukoulis, S. Linden, "Negative-index metamaterial at 780 nm wavelength," http://arxiv.org/abs/physics/0607135, (2006).

## 1. Introduction

A thin film of a nanostructured metamaterial with physical thickness $\delta$ can be characterized through its spectra to have an effective refractive index $n = n' + \iota n''$ and an effective impedance $\eta = \eta' + \iota \eta''$. In addition, along with its effective $n$ and $\eta$, the layer can be characterized by its effective permittivity $\varepsilon = \varepsilon' + \iota \varepsilon''$ and permeability $\mu = \mu' + \iota \mu''$, obtained as $\varepsilon = n/\eta$, and $\mu = n\eta$. The values of $n$ and $\eta$ (or $\varepsilon$ and $\mu$) of the equivalent homogenized layer of thickness $\delta$ are chosen to reproduce the complex values of far-field reflectance and transmittance due to a given film. Using this technique [1,2], the complex values of the transmitted and reflected fields are obtained either from optical experiments or simulations.

Optical negative index metamaterials (NIMs), also known as left-handed materials, are artificially engineered metal-dielectric composites that exhibit $n' < 0$ within a certain range of wavelengths. In addition, a magnetic resonant behavior should be observed in NIMs at this range. The magnetic resonance in any optical NIM is always required to make the real part of the effective refractive index negative, either through the strong (sufficient) condition $\mu' < 0$ and $\varepsilon' < 0$, or through a more general necessary condition $\varepsilon'\mu'' + \mu'\varepsilon'' < 0$, which is valid for a passive medium. The general condition strictly implies that there is no negative refraction effect in a passive metamaterial with $\mu = 1 + 0\iota$. Nonetheless, the effect is also achievable for $\mu' > 0$ provided that only $\varepsilon' < 0$ in $\varepsilon'\mu'' + \mu'\varepsilon''$, and $|\varepsilon'\mu| > |\mu'\varepsilon|$. In the latter case, substantial 'magnetic' losses are necessary along with a dominant metal content in the structure. We note that a ratio of $|n'/n''|$ is often taken as a figure of merit (FOM) for NIM performance, since low-loss NIMs are desirable. The FOM can be written as $|n'/n''| = |\varepsilon'|\mu| + |\varepsilon|\mu'|/|\varepsilon''|\mu| + |\varepsilon|\mu''|$, indicating that a 'double-negative' NIM ($\mu' < 0$, $\varepsilon' < 0$) will have a much better FOM than a NIM layer with $\mu' > 0$. Therefore, metal-dielectric composites with a negative effective permeability are essential for further development of low-loss optical NIMs and their applications.

Recent computational results [3,4] have demonstrated that pairs of thin silver strips separated by a dielectric spacer could offer an easy way of achieving negative magnetism by coupling near-field modes. This approach generalizes the original idea of a diamagnetic response (and negative n) in pairs of rods [5]. It has also been shown that the magnetic resonance of a periodic array of coupled silver strips with sub-wavelength periodicity is always accompanied by an electric anti-resonance; this is the periodicity effect and it does not occur for the case of an isolated strip pair of the same structural dimensions and materials. Recent studies [3,5,7] show that the destructive effect of the electric anti-resonance, which tends to make $\varepsilon'$ (as well as $n'$) positive, can be straightforwardly compensated by adding background metallic elements (non-resonant strips or wires, homogeneous or inhomogeneous films), and that it is mostly the negative effective permeability that holds the key for advancing the design of low-loss NIMs in optics. Previously reported important results with different periodic metal-dielectric composites have already been obtained in the terahertz and subsequently in the infrared spectral ranges [7-11]. (Herein, we are not considering a recent report on negative magnetism and negative refractive index in the green light range [12] because the following careful studies performed by two independent groups show that both claims do not have grounds [13].)

This paper deals with the experimental observation of a negative permeability in the visible range in a periodic array of pairs of thin silver strips. For our study, two samples (denoted as Sample A and B) with slightly different geometries have been fabricated. A negative effective permeability has been retrieved using numerical simulations; the results are in good agreement with the transmission and reflection spectra obtained from optical measurements for each sample. The value of $\mu'$ is about −1 in Sample A and about −1.7 in Sample B at the wavelengths of 770 nm and 725 nm, respectively.

In addition to the predicted negative permeability and electric anti-resonances, abnormal anisotropic losses, occurring near the resonances and resulting from imperfections in the fabricated NIM structures are also observed experimentally. Thus, we demonstrate a negative magnetic response from a periodic optical material for visible (red) light and discuss new challenges due to significantly increased losses in the optical properties of "imperfect" thin silver strips observed at the resonances.

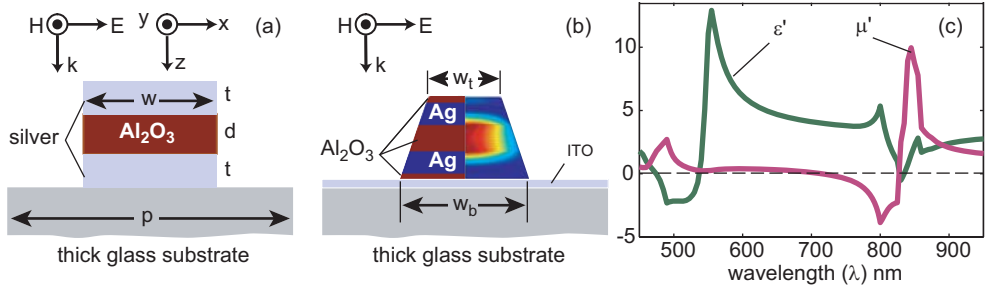

Fig. 1. (a) Ideal unit cell for the array of coupled silver nano-strips of width $w$ which are separated by a strip of alumina with the same width; here $t$ is the thickness of both strips, and $d$ is the thickness of the alumina spacer. The strips are infinite in $y$ direction and periodic in $x$ direction with period $p$. (b) The actual cross-section of samples obtained after fabrication (left half). Right half shows the map of magnetic field enhanced at the magnetic resonance. (c) The real part of permeability and permittivity shown for the cell with $w$ = 140 nm, $t$ = 35 nm, $d$ = 40 nm, and $p$ = 300 nm. The optical constants of bulk silver [15] are taken for the strips. The refractive index of the glass substrate is 1.52.

Figure 1(a) shows an initial ideal elementary cell that we have used to optimize a negative permeability sub-wavelength grating. It consists of a pair of thin silver strips (with thickness $t$ and width $w$). The strips are separated by an alumina spacer with thickness $d$, width $w$, and a refractive index of $n = 1.62 + 0\iota$. The sub-wavelength lattice constant of the grating is $p$. The structure has been optimized using custom code based on the spatial harmonic analysis (SHA) approach [14] with additional fabrication constraints pertinent to electron beam lithography. In the optimal structure the periodicity $p$ was chosen to be 300 nm with $t$ = 35 nm, $d$ = 40 nm, and $w$ = 140 nm, and isotropic bulk optical properties of silver have been taken from [15]. In the resonant (TM) polarization the magnetic field is aligned with the largest dimension – the infinite length of the strips. Only one component of the magnetic field should ideally be present in this case. In the non-resonant (TE) polarization the single component of the electric field is aligned with the strip length, giving no resonant effects. In such an ideal sub-wavelength grating of Fig. 1(a), a relatively wide negative magnetic response exists in the TM regime and it extends from a wavelength of 720 nm to 825 nm, as shown in Fig. 1(c). A sharp electric resonance behavior is also demonstrated for the structure in TM mode around 500 nm. The electric resonance introduces a magnetic anti-resonance response within their common wavelength range. A reversed effect is observed at the magnetic resonance, where the electric anti-resonance is now present. The presence of anti-resonance makes it difficult to overlap the magnetic and electric resonances, since as the electric and magnetic resonances get closer to each other, the anti-resonances increase in strength, resulting in the damping of the resonances [3].

Fig. 1(b) shows a cross-section of the structure adjusted relative to the ideal structure of Fig. 1(a) in order to reflect fabrication realities. Unavoidable imperfections of the fabrication procedure result in a trapezoidal shape of the stacked strips Therefore, in contrast to the ideal structure, in the actual cross-section the top width $w_t$ is smaller than the bottom width $w_b$. In addition, two thin 10-nm layers of alumina are added, one between the lower silver strip and the substrate and the second on top of the structure. Both additional alumina layers appear to be necessary for the stable fabrication of samples. Electron beam lithography techniques have been used to fabricate the samples.

First, the geometry of the periodic thin silver strips was defined on a glass substrate initially coated with a 15-nm film of indium-tin-oxide (ITO) by use of an electron beam writer. Then, electron beam evaporation was applied to produce a stack of lamellar films.

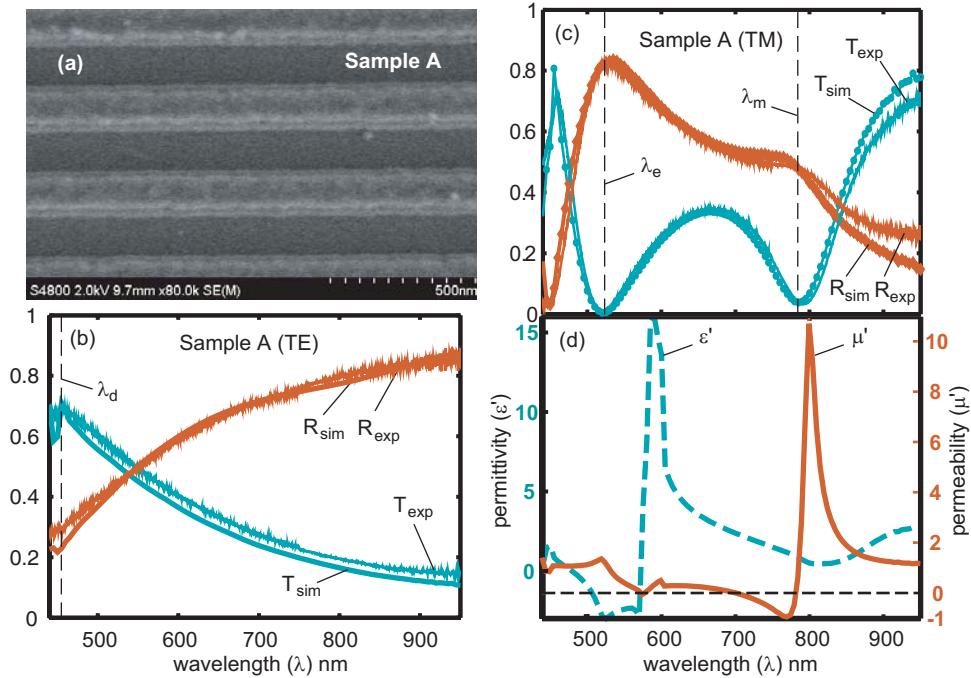

Fig. 2. (a) FE SEM picture of the periodic array of coupled silver strips (Sample A). (b) Transmission and reflection spectra of Sample A measured at normal incidence with TE polarization, here $\lambda_d$ is the diffraction threshold. The experimental spectra are compared to the results of numerical modeling. The optical constants of silver strips are taken from the experimental data for bulk silver [15] (c) Transmission and reflection spectra of Sample A at normal incidence with TM polarization compared to spectra obtained from simulations. In this case, $\varepsilon''$ of the silver strips was adjusted to match excessive losses. (d) The real part of the effective permeability ($\mu'$) and effective permittivity ($\varepsilon'$).

Finally, a lift-off process was performed to obtain the desired silver strips. The projected serial structure of the films from the ITO-coated glass was: Sample A, 10-nm alumina, 30-nm silver, 40-nm alumina, 30-nm silver, 10-nm alumina; Sample B, 10-nm alumina, 35-nm silver, 40-nm alumina, 35-nm silver, 10-nm alumina. As an example of the fabricated structure, a FE SEM image of Sample A is shown in Fig. 2(a).

To test the fabricated samples, we measured the transmission and reflection spectra of the samples with an ultra-stable tungsten lamp (B&W TEK BPS100). The spectral range of the lamp covers the entire visible and near-infrared optical band. A Glan Taylor prism was placed at the output of the broadband lamp to select the light with desired linear polarization. The signal transmitted (or reflected) from the sample was introduced into a spectrograph (Acton SpectraPro 300i) and eventually collected by a liquid nitrogen cooled CCD-array detector. The transmission and reflection spectra were normalized to a bare substrate and a calibrated silver mirror, respectively. In the TE regime the electric field of the incident light was linearly polarized parallel to the length of silver strips, while in TM mode the electric field was rotated 90 degrees relative to TE case. For example, Figs. 2(b) and 2(c) show transmission and reflection spectra obtained from the optical measurements of Sample A for TE and TM polarizations at normal incidence.

In our simulations with a commercial finite element software (more suitable for modeling the exact geometrical details of the structure than the SHA code), an incident plane wave source was placed at the source end of the computational domain, and the transmitted and reflected field amplitudes were monitored at two points located inside the domain several wavelength away from the film under test.

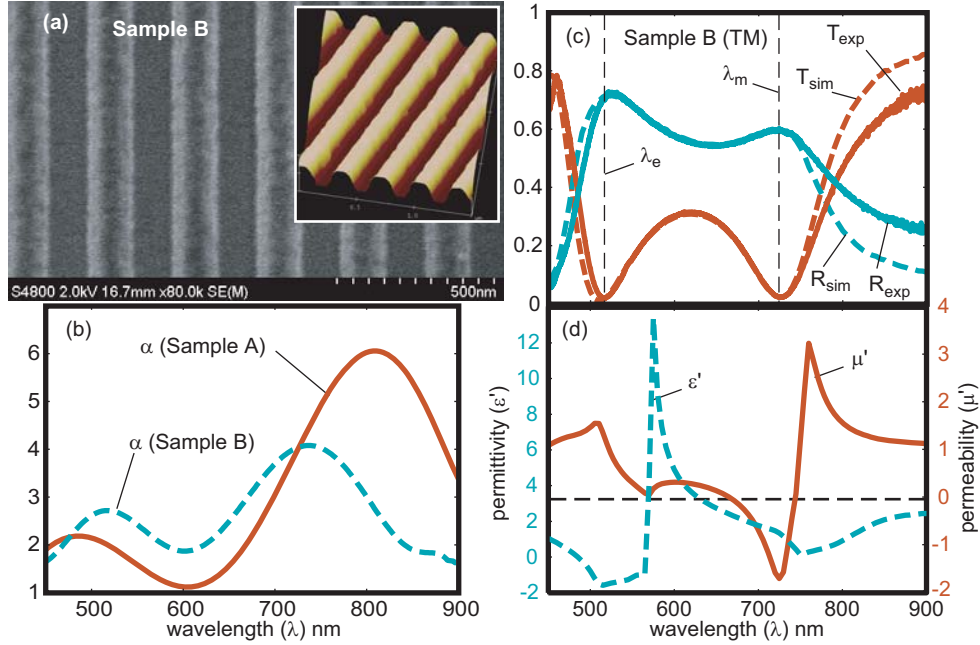

Fig. 3. (a) FE SEM picture of Sample B. (The inset shows an AFM image of the sample.) (b) Comparison of the loss-adjustment factor $\alpha$ obtained for Samples A and B. Sample A demonstrates more excessive loss in comparison to bulk metal [15] and Sample B. (c) Transmission and reflection spectra of Sample B at normal incidence with TM polarization compared to spectra obtained from simulations. In this case, $\varepsilon''$ of the silver strips was adjusted to match excessive losses. (d) The real part of the effective permeability ($\mu'$) and effective permittivity ($\varepsilon'$). Minimal $\mu'$ of $-1.7$ is obtained at 725 nm.

The optical constants of silver used in the ideal model have been initially taken from the experimental data [15] to obtain the optimal structure; as a result, a substantially negative $\mu'$ has been obtained, as shown in Fig. 1(c). In the ideal situation, both the experimental and simulation setup would allow for an adequate match of the reflection and transmission spectra of light in both polarizations. The transmission spectrum with TE polarization is shown in Fig. 2(c). In this case, as expected, both spectra match well over a broad range of measured wavelengths. The measured spectra display a moderate non-resonant wavelength dependence and low, almost constant, absorption; transmission falls off closer to the higher wavelengths. The relaxed wavelength dependence is attributed to non-resonant behavior of the metallic strips diluted with the alumina spacer in a layer that could be adequately described by an effective medium theory. Good matching occurs almost everywhere, provided that the light in the measured and simulated structures has a single propagating mode (no diffraction). Beyond 950 nm, the signal-to-noise ratio worsens, making signal detection difficult at the output.

A useful feature of the optical TE measurements is that the spectral position of the diffraction threshold, shown in Fig. 2(c) as $\lambda_d$, is a direct indicator of the true effective periodicity observed in the experiment. Indeed, as long as the refractive index of the substrate, $n_s$, is known almost exactly, a simple and accurate measurement of the actual period can be obtained from $p = \lambda_d / n_s$.

Similar to the TE mode, the TM polarization gives useful data for measuring the actual geometry of the samples obtained after fabrication. Specifically, the spectral position of the electric and magnetic resonances $\lambda_e$ and $\lambda_m$ shown in Fig. 2(c) are both very sensitive to the thickness and width of the metallic strips. Good agreement with the experimental spectra of Figs. 2(c) and 2(b) has been achieved by varying these parameters within realistic fabrication tolerances. Thus, the silver strips used in the simulation of Sample A were 4 nm thinner and 8 nm wider than initial estimates from FE SEM images. Additional examinations of FE SEM images taken in a number of areas over the sample are consistent with this result. As expected, spectroscopic measurements appeared to be more accurate for the dimensions that are critical in the resonant regimes, giving a better result than estimates based on FE SEM images. In addition, clear-cut optical measurements of transmission and reflection spectra with both polarizations incorporate and level out imperfections inevitable in the fabrication process. It is also important to note that the deviations from the initially planned thickness agree well with roughness estimates taken at the strip surface. Thus, in Sample A, where the surface roughness is within 5 – 6 nm, the actual thickness of the strips obtained by matching numerical simulations to spectroscopic measurements is about 26 nm versus 30-nm thickness planned for this design. In accord with this, the strips of Sample B fabricated with a lower deposition rate and demonstrating much less roughness (about 2 – 3 nm) do not show any discernable deviation from a planned thickness of 35 nm. The dimensions of the samples taken in the simulations are: Sample A, $t = 26$ nm, $d = 48$ nm, $w_t = 94$ nm, and $w_b = 174$ nm; Sample B, $t = 35$ nm, $d = 40$ nm, $w_t = 85$ nm, and $w_b = 160$ nm.

In contrast to the TE case, where the optical constants of the bulk silver are used to describe the optical behavior of silver strips, the TM polarization reveals substantial discrepancies in the optical properties (specifically, in losses) of nano-structured silver strips vs. the data shown for bulk silver in [15]. Specifically, the experimental loss is larger than that obtained through simulations. This enhanced loss is due to the roughness features and other imperfections in the fabricated structures. We model contributions from these imperfections through a wavelength dependant adjustment factor ($\alpha$), such that the permittivity of silver is given by, $\varepsilon = \varepsilon' + \iota\alpha\varepsilon''$.

The simulated absorbance was matched with the experimental absorbance by incorporating a suitable $\alpha$. This deviation appears to be significant at the electric and magnetic resonances of the fabricated samples, whereas outside the resonances $\alpha$ is close to 1. Since at the resonances the spatial spectral content of the near-field is radically increasing, we conclude that the near-field modes strongly interact with the inhomogeneities of the silver structures, resulting in the observed additional losses. The dominant symmetric modes of the electric resonance are less sensitive to the imperfections of the strips so that relatively smaller adjustments of $\varepsilon''$ in silver are necessary around the electric resonance. At the same time, the steeper gradients of the asymmetric modes dominant at the magnetic resonance are much more sensitive to structural defects and surface roughness of the strips so that a substantial adjustment of losses is required in this case.

Computed results of the transmission and reflection spectra using the adjusted (imaginary part of) permittivity of bulk silver are shown in Figs. 2(c) and 3(c); the used adjustment factor $\alpha$ is shown in Fig. 3(b). In both samples the loss adjustment is significant for the electric resonance and especially large for the magnetic resonance. The adjustment is almost one between the two resonances, and it reaches its upper level of about six for Sample A and almost seven for Sample B at the magnetic resonances. The additional losses in the resonant regimes of the strips are very large. Unfortunately, these losses diminish the negative magnetic response over the entire range of the magnetic resonance.

Away from the electric and magnetic resonances at both polarizations, the presence of small roughness features on the metal strips does not have much effect so that almost no adjustment to ε″ is needed and $\alpha \approx 1$. In contrast, for the TM polarization near the magnetic resonance, large values of $\alpha$ have to be used. This is because the asymmetric currents in the silver strips (causing the magnetic response) are "closed" by the *vertical* displacement field which induces local surface plasmons (LSPs) in the roughness features on the surface of the metal strips, causing the increased losses. The role of LSPs is somewhat less for the electric resonance in TM polarization because this resonance is associated with the *horizontally* (x) oriented E field, which should be continuous at the metal-dielectric interface.

A similar correction for $\varepsilon''$ of silver in an optical metamaterial has been recently implemented by using a wavelength-independent factor ($\alpha = 3$) [16]. Fig. 4 illustrates this approach, where $\alpha = 3$ is also used. Fig. 4(a) shows that the constant $\alpha$ demonstrates somewhat better conformity with the T and R spectra than shown in [16], but still worse than the agreement achieved in Fig. 3(c) using the wavelength-dependent factor of Fig. 3(b). In addition, as shown in Figs. 4(a) and (b) underestimated losses tend to make $\mu'$ artificially more negative.

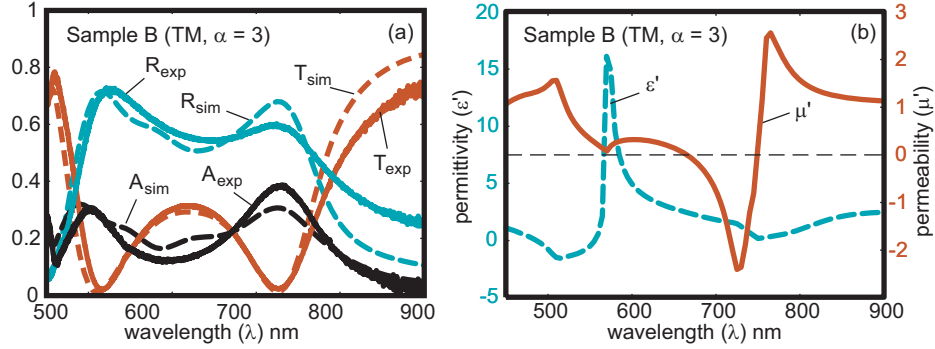

Fig. 4. (a) Transmission and reflection spectra of Sample B at normal incidence with TM polarization compared to spectra obtained from simulations; here $\varepsilon''$ of silver was adjusted using a factor of $\alpha = 3$. (b) The real part of the effective values of $\mu'$ and $\varepsilon'$ obtained using the wavelength-independent factor, $\alpha = 3$.

The ideal spectra of $\mu'$ computed without any adjustment to the permittivity of bulk silver and the values of $\mu'$ calculated with the adjustment factor of Fig. 3(b) have been also compared. Relative to the ideal metal, negative magnetism is reduced by a factor of 7.8 in Sample A and only by a factor of 2.4 in Sample B. By measuring and comparing the transmission and reflection of other samples with different fabrication conditions, the average loss adjustment factor of the anisotropic permittivity of silver is estimated to range from four to six over the entire magnetic resonance. Although the actual range of the adjustment factor $\alpha$ depends greatly both on the initial design of the magnetically resonant structure and its fabrication conditions, these estimates provide approximate margins for the adjusted values of losses in resonant nano-structured silver structures in the visible light range. Most importantly, this comparison clearly indicates that by fabricating strips with minimized roughness features the magnetic response can be dramatically increased.

In summary, we designed, fabricated and measured two samples with negative permeability of –1 and –1.7 at 770 nm and 725 nm, respectively. Detailed numerical models of the samples have been used to validate the measurements, where an adjusted wavelength-dependent $\varepsilon''$ is used in silver at the resonances. The results of the numerical modeling are compared with the experimental data. The good agreement between experiment and theory achieved for two different polarizations in a wide range of wavelength is an excellent confirmation of the validity of the model, hence confirming the existence of negative permeability at red light. We also discuss the difficulties in automated optimization of NIMs due to substantial deviations from the properties (specifically, losses) of bulk metal observed experimentally in the resonant plasmonic elements of our nano-structured samples. These deviations would require constant feedback from fabrication and measurements for obtaining double-negative NIMs in visible range. The observed results also indicate that by further improving the fabrication so that the imperfections causing the increased losses are minimized one can obtain a much stronger negative magnetic response.

This work was supported in part by ARO grant W911NF-04-1-0350, NSF-PREM Grant #DMR-0611430 and by ARO-MURI award 50342-PH-MUR.